\documentclass[a4paper,11pt]{article}

\usepackage[a4paper,left=2.73cm,right=2.7cm,top=3cm,bottom=3.5cm]{geometry}

\usepackage{amsfonts,amsmath,amssymb,tabstackengine}
\usepackage[usenames, dvipsnames]{color}
\usepackage{array}
\usepackage{tikz}

\usepackage{color, colortbl}
\definecolor{Gray}{gray}{0.95}

\usepackage[colorlinks=true,linktocpage=true,linkcolor=blue,citecolor=blue]{hyperref}
\usepackage{graphicx}

\addtocontents{toc}{\protect\setcounter{tocdepth}{3}}
\numberwithin{equation}{section}

\begin{document}

\begin{titlepage}

\thispagestyle{empty}

\begin{center}

{\LARGE \textbf{Flat deformations of type IIB S-folds}}

\vspace{40pt}
		
{\large \bf Adolfo Guarino}$\,^{a, b}$\,\,\,\,  \large{and}  \,\,\,\, {\large \bf Colin Sterckx}$\,^{c, a}$ 
		
\vspace{25pt}
		
$^a$\,{\normalsize  
Departamento de F\'isica, Universidad de Oviedo,\\
Avda. Federico Garc\'ia Lorca 18, 33007 Oviedo, Spain.}
\\[7mm]

$^b$\,{\normalsize  
Instituto Universitario de Ciencias y Tecnolog\'ias Espaciales de Asturias (ICTEA) \\
Calle de la Independencia 13, 33004 Oviedo, Spain.}
\\[7mm]

$^c$\,{\normalsize  
Universit\'e Libre de Bruxelles (ULB) and International Solvay Institutes,\\
Service  de Physique Th\'eorique et Math\'ematique, \\
Campus de la Plaine, CP 231, B-1050, Brussels, Belgium.}
\\[10mm]

\texttt{adolfo.guarino@uniovi.es} \,\, , \,\, \texttt{colin.sterckx@ulb.ac.be}

\vspace{20pt}

%\today

\vspace{20pt}
				
\abstract{
\noindent 

Type IIB S-folds of the form $\textrm{AdS}_{4} \times \textrm{S}^1 \times \textrm{S}^5$ have been shown to contain axion-like deformations parameterising flat directions in the 4D scalar potential and corresponding to marginal deformations of the dual S-fold CFT's. In this note we present a group-theoretical characterisation of such flat deformations and provide a 5D interpretation thereof in terms of $\mathfrak{so}(6)$-valued duality twists inducing a class of Cremmer--Scherk--Schwarz flat gaugings in a reduction from 5D to 4D. In this manner we establish the existence of two flat deformations for the $\mathcal{N}=4$ and $\textrm{SO}(4)$ symmetric S-fold causing a symmetry breaking down to its $\textrm{U}(1)^2$ Cartan subgroup. The result is a new two-parameter family of non-supersymmetric S-folds which are perturbatively stable at the lower-dimensional supergravity level, thus providing the first examples of such type IIB backgrounds.

}

\end{center}

\end{titlepage}

\tableofcontents

\hrulefill
\vspace{10pt}

\section{Motivation and summary of results}
\label{sec:intro}

S-fold backgrounds of type IIB supergravity of the form $\textrm{AdS}_{4} \times \textrm{S}^1  \times \textrm{S}^5$ \cite{Inverso:2016eet} have recently received much attention due to their holographic interpretation as new strongly coupled three-dimensional conformal field theories (CFT’s) on a localised interface of super-Yang--Mills (SYM) \cite{Assel:2018vtq}. On the gravity side, S-folds have been obtained in two complementary manners: $i)$ as AdS$_{4}$ vacua of an effective four-dimensional dyonically-gauged supergravity with gauge group $[\textrm{SO}(1,1) \times \textrm{SO}(6)] \ltimes \mathbb{R}^{12}$ and electromagnetic deformation parameter $\,c\,$ \cite{Inverso:2016eet,Guarino:2019oct,Guarino:2020gfe}. $ii)$ as limiting Janus solutions of an effective five-dimensional gauged supergravity with gauge group $\textrm{SO}(6)$ \cite{Bobev:2019jbi,Bobev:2020fon,Arav:2021tpk}. Both approaches rely on the consistency of the reduction of type IIB supergravity on $\,\textrm{S}^5\,$ \cite{Baguet:2015sma} as well as on $\, \textrm{S}^1  \times \textrm{S}^5\,$ \cite{Inverso:2016eet}, respectively, the latter incorporating an additional non-trivial SL(2)$_{\textrm{IIB}}$ hyperbolic monodromy \cite{Hull:2003kr} in the reduction ansatz along $\textrm{S}^1$ that depends on the electromagnetic parameter $\,c\,$ and is responsible for the $\textrm{SO}(1,1)$ factor in the gauge group.

After the $\,\mathcal{N}=4\,$ and $\,\textrm{SO}(4)\,$ symmetric S-fold originally obtained in \cite{Gallerati:2014xra} using the 4D approach, various multi-parametric families of S-folds were obtained in \cite{Guarino:2020gfe} preserving different amounts of supersymmetry (${\mathcal{N}=0,1,2}$) and of residual symmetry. These families generically depend on a set of axion-like parameters, denoted $\,\chi$'s in \cite{Guarino:2020gfe}, which correspond to flat directions of the scalar potential of the four-dimensional $[\textrm{SO}(1,1) \times \textrm{SO}(6)] \ltimes \mathbb{R}^{12}$ gauged supergravity, and thus to exactly marginal deformations in the holographic S-fold CFT duals. The family of $\,\mathcal{N}=0\,$ S-folds depends on three such parameters $\,\chi_{1,2,3}\,$ which specify the matrix
\begin{equation} 
\label{chi_matrix}
\chi^{ij} = \left(
\begin{array}{cccccc}
0 & \chi_{1} & & & & \\
-\chi_{1} & 0 & & & & \\
 & & 0 & \chi_{2} & & \\
 &  & -\chi_{2} & 0 & & \\
 & & & & 0 & \chi_{3}  \\
 & & & & -\chi_{3} & 0
\end{array}
\right) \in \mathfrak{so}(2)^3  \subset \mathfrak{so}(6) \ , 
\end{equation}
that controls the residual symmetry group of the corresponding S-fold solution. This group ranges from $\,\textrm{U}(1)^3\,$ at generic values of $\,\chi_{1,2,3}\,$ to $\,\textrm{SO}(6) \sim \textrm{SU}(4)\,$ at the special case $\,\chi^{ij}=0\,$. The family of $\,\mathcal{N}=1\,$ S-folds depends on the axions in (\ref{chi_matrix}) which are this time subject to the condition $\,{\chi_{1} + \chi_{2} + \chi_{3}=0}\,$. The residual symmetry group of the S-fold solution ranges from $\,\textrm{U}(1)^2\,$ at generic values of the axions to $\,\textrm{SU}(3)\,$ at the special case $\,\chi^{ij}=0\,$. Lastly, the family of $\,\mathcal{N}=2\,$ S-folds depends on the axions in (\ref{chi_matrix}) subject now to the identifications $\,\chi \equiv \chi_{1} = - \chi_{3}\,$ and $\,\chi_{2} = 0\,$. The symmetry group of the S-fold solution ranges from $\,\textrm{U}(1)^2\,$ at a generic value of the axion $\,\chi\,$ to $\,\textrm{SU}(2) \times \textrm{U}(1)\,$ at the special case $\,\chi^{ij}=0\,$.\footnote{The study and characterisation of the entire conformal manifold of $\,\mathcal{N}=2\,$ S-fold CFT's has also been the focus of recent works \cite{Arav:2021gra,Bobev:2021yya}. The $\chi$-family of $\,\mathcal{N}=2\,$ S-folds was generalised to a larger two-parameter $(\chi,\varphi)$-family of S-folds accommodating the original $\,\mathcal{N}=4\,$ and $\,\textrm{SO}(4)\,$ symmetric S-fold as a special case \cite{Bobev:2021yya}.} One quickly realises that turning on the axions $\,\chi^{ij}\,$ induces a sequential symmetry breaking from the largest possible symmetry at $\,\chi^{ij}=0\,$ down to its Cartan subgroup.

The aim of this note is to disclose the universal character of the axion-like parameters $\,\chi^{ij} \in \mathfrak{so}(6)\,$ in the families of type IIB S-folds, and to understand why they give rise to flat directions in the scalar potential and, therefore, to potential marginal deformations in the dual S-fold CFT's. To this end we will take advantage of the $\,\textrm{E}_{7(7)}/\textrm{SU}(8)\,$ coset structure of the scalar manifold in the $\,\mathcal{N}=8\,$ supergravity multiplet. Being an homogeneous space, any two scalar field configurations are connected by an $\,\textrm{E}_{7(7)}\,$ transformation. Using the formally $\textrm{E}_{7(7)}$-covariant formulation of the maximal 4D gauged supergravities provided by the embedding tensor formalism \cite{deWit:2007mt}, it is always possible to map an S-fold solution with non-vanishing axions $\,\chi^{ij} \neq 0\,$ in the gauged supergravity with gauge group $\,\textrm{G}=[\textrm{SO}(1,1) \times \textrm{SO}(6)] \ltimes \mathbb{R}^{12}\,$ to a new S-fold solution this time with vanishing axions $\,\chi^{ij} = 0\,$. However, $\textrm{E}_{7(7)}$-covariance requires having to act on the embedding tensor $\,\Theta\,$ itself which fully specifies the gauging and the couplings in the supergravity Lagrangian. Then, bringing the axions to $\,\chi^{ij}=0\,$ implies having to change the original gauged supergravity to a new one based on a new gauge group $\,\widetilde{\textrm{G}}\,$ specified by a new embedding tensor $\,\widetilde{\Theta}\,$. This method of mapping solutions belonging to different theories has already been exploited in the past to chart the landscape of AdS$_{4}$ vacua in half-maximal \cite{Dibitetto:2011gm} as well as maximal \cite{DallAgata:2011aa,Borghese:2012qm,Borghese:2013dja} supergravities in four dimensions.

As we will show in this note, and upon application of the above procedure, the new theory will be a superposition of the original $[\textrm{SO}(1,1) \times \textrm{SO}(6)] \ltimes \mathbb{R}^{12}$ dyonically-gauged supergravity and a class of Cremmer--Scherk--Schwarz (CSS) flat gaugings \cite{Cremmer:1979uq}. More concretely, three out of the four mass parameters in a CSS gauging are directly mapped into the axion-like parameters $\,\chi^{ij} \in \mathfrak{so}(6)\,$ in (\ref{chi_matrix}), \textit{i.e.} $\,{m_{1,2,3}=\chi_{1,2,3}}\,$. On the contrary, being associated to a non-compact $\,\mathfrak{so}(1,1)\,$ duality twist \cite{Inverso:2016eet}, the electromagnetic parameter $\,c\,$ is not part of the CSS gauging. The fourth mass parameter in a CSS gauging is thus absent, \textit{i.e.} $\,{m_{4}=0}\,$, and we arrive at the general structure
\begin{equation}
\label{Theta_splitting}
\widetilde{\Theta} \,\,=\,\, \Theta^{[\textrm{SO}(1,1) \times \textrm{SO}(6)] \ltimes \mathbb{R}^{12}} \,\,+\,\, (\delta\Theta)^{\textrm{CSS}}(\chi_{1} \,,\, \chi_{2} \,,\, \chi_{3} \,,\, 0) \ .
\end{equation}
We will show the connection between the axion-like parameters $\,\chi^{ij} \in \mathfrak{so}(6)\,$ and the existence of flat directions in the scalar potential of the original $\,[\textrm{SO}(1,1) \times \textrm{SO}(6)] \ltimes \mathbb{R}^{12}\,$ gauged supergravity. We will do it by proving that the contribution of $\,(\delta\Theta)^{\textrm{CSS}}\,$ in (\ref{Theta_splitting}) to the scalar potential induced by $\,\widetilde{\Theta}\,$ identically vanishes when performing a group-theoretical truncation of the scalar field content to the $\textrm{G}^{\chi}_{0}$-invariant sector, where $\textrm{G}^{\chi}_{0}\,$ is a Cartan subgroup of $\,\textrm{G}_{0} \subset \textrm{SO}(6)\,$ chosen to commute with $\chi^{ij}$ and $\,\textrm{G}_{0}\,$ is the largest symmetry group of the axion-vanishing solution within a family of S-folds.

\begin{figure}[t]
\begin{center}
\includegraphics[width=0.68\textwidth]{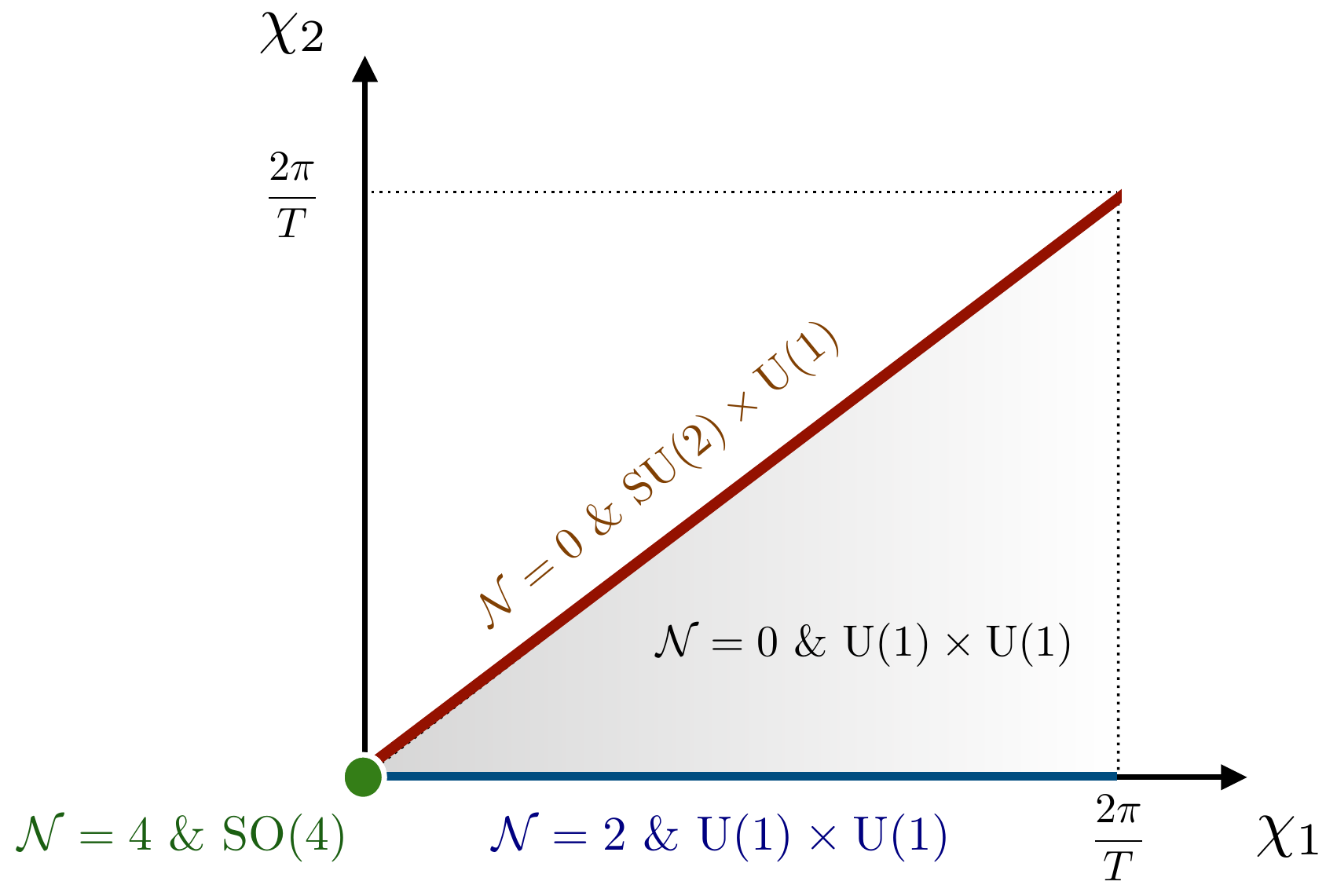}
\caption{Two-dimensional parameter space $\,(\chi_{1},\chi_{2})\,$ of S-fold solutions induced by axion-like flat deformations of the $\,\mathcal{N}=4\,$ and $\,\textrm{SO}(4)\,$ symmetric S-fold (green dot). The blue and red lines correspond to special choices of parameters $\,\chi_{1,2}\,$ and produce supersymmetry (blue, $\,\chi_{2}=0\,$) and residual symmetry (red, $\,\chi_{1}=\chi_{2}\,$) enhancements. These two special lines define the boundary of the parameter space.}
\label{fig:diagram}
\end{center}
\end{figure}

As an illustration of the method, we establish the existence of two flat deformations $\,\chi_{1,2}\,$ of the $\,\mathcal{N}=4\,$ and $\,\textrm{SO}(4)\,$ symmetric S-fold in \cite{Inverso:2016eet} which lie outside the $\,\textrm{SO}(4)$-invariant sector of the theory. These two flat deformations are responsible for the breaking of the $\,\textrm{SO}(4)\,$ symmetry of the $\,\mathcal{N}=4\,$ S-fold down to its $\,{\textrm{U}(1) \times \textrm{U}(1)}\,$ Cartan subgroup. Setting one of the deformation parameters to zero, \textit{e.g.} $\,\chi_{2}=0\,$, a known class of $\,\mathcal{N}=2\,$ S-folds with $\,\textrm{U}(1)^2\,$ symmetry is recovered\footnote{It corresponds to setting $\,(\chi,\varphi)=(\frac{\chi_{1}}{\sqrt{2}},1)\,$ in the general two-parameter family of $\,\mathcal{N}=2\,$ and $\,\textrm{U}(1)^2\,$ symmetric S-folds of \cite{Bobev:2021yya}.\label{footnote_N2_manifold}}. (Anti-)identifying the two parameters yields a one-parameter family of non-supersymmetric, yet perturbatively stable, S-folds with $\,\textrm{SU}(2) \times \textrm{U}(1)\,$ symmetry.\footnote{The existence of non-supersymmetric AdS$_{4}$ critical points in the $\,[\textrm{SO}(1,1) \times \textrm{SO}(6)] \ltimes \mathbb{R}^{12}\,$ gauged maximal supergravity has already been established in \cite{DallAgata:2011aa}, and further investigated in \cite{Guarino:2019oct,Guarino:2020gfe,Bobev:2021yya} by looking at the $\,\mathbb{Z}_{2}^{3}$-invariant sector of the theory. However all such non-supersymmetric AdS$_{4}$ extrema of the scalar potential exhibit BF instabilities \cite{Breitenlohner:1982bm}.} Lastly, a generic choice of the two parameters $\,\chi_{1,2}\,$ produces non-supersymmetric S-folds with $\,\textrm{U}(1)^2\,$ symmetry which are also perturbatively stable at the lower-dimensional supergravity level. The invariance of the entire setup under the reflection $\,\chi_{i} \rightarrow - \chi_{i}\,$ and the exchange $\,\chi_{1} \leftrightarrow \chi_{2}\,$ reduces the parameter space of the S-fold solutions to the octant depicted in Figure~\ref{fig:diagram}.

On the other hand, the geometric interpretation of the axion-like parameter $\,\chi\,$ in the family of $\,{\mathcal{N}=2}\,$ \mbox{S-folds} was presented in \cite{Giambrone:2021zvp}. By looking at the ten-dimensional uplift of the entire family of S-folds, the parameter was shown to be compact, \textit{i.e.} $\,\chi \in [0,2 \pi/T)\,$, and to induce a non-trivial $\chi$-dependent fibration of $\,\textrm{S}^5\,$ over the $T$-periodic $\,\textrm{S}^1\,$. This interpretation holds for the parameters in (\ref{chi_matrix}) determining the other families of $\,\mathcal{N}=0,1\,$ S-folds: they are also compact and induce a non-trivial monodromy on the internal $\,\textrm{S}^5\,$ when moving around $\,\textrm{S}^1\,$. The specific monodromy element $\,h(\chi_{1,2,3})\,$ controls the patterns of symmetry breaking as classified by the mapping torus $\,T_h(\textrm{S}^5)\,$ \cite{Guarino:2021kyp}. It would then be interesting to uplift the flat deformations $\,\chi_{1,2}\,$ of the $\,\mathcal{N}=4\,$ and $\,\textrm{SO}(4)\,$ symmetric S-fold to ten dimensions using $\,\textrm{E}_{7(7)}$-covariant Exceptional Field Theory techniques along the lines of \cite{Inverso:2016eet,Guarino:2019oct,Guarino:2020gfe}. It is a reasonable expectation that the axions $\,\chi_{1,2}\,$ can be interpreted in terms of non-trivial monodromies on $\,\textrm{S}^5\,$ when moving around $\,\textrm{S}^1$, and that the patterns of symmetry breaking they induce are still controlled by the mapping torus. This would render the axions compact, \textit{i.e.} $\,\chi_{1,2} \in [0,2 \pi/T)\,$,  and so the parameter space they span (shadow region in Figure~\ref{fig:diagram}).

The note is structured as follows. In section~\ref{sec:4D_sugra} we introduce the $[\textrm{SO}(1,1) \times \textrm{SO}(6)] \ltimes \mathbb{R}^{12}$ dyonically-gauged maximal 4D supergravity and provide a characterisation of its consistent flat deformations as induced by a set of axions $\,\chi^{ij} \neq 0\,$. To illustrate the general method, we work out explicitly the set of flat deformations of the $\,\mathcal{N}=4\,$ S-fold with $\,\textrm{SO}(4)\,$ symmetry and find new classes of non-supersymmetric S-fold solutions. In section~\ref{sec:5D_CSS} we provide a 5D interpretation of the flat deformations in terms of duality twists inducing CSS gaugings in a reduction from 5D to 4D. We also present a group-theoretical unification of the axion-like deformations $\,\chi^{ij}\,$ and the electromagnetic parameter $\,c\,$, emphasising their similarities and main differences. Two appendices accompany the note. Appendix~\ref{app:E7-conventions} collects our conventions on the $\,\mathfrak{e}_{7(7)}\,$ algebra. Appendix~\ref{app:Proof} contains a proof of the result (\ref{V=V=V}) used in the main text.

\section{The maximal $[\textrm{SO}(1,1) \times \textrm{SO}(6)] \ltimes \mathbb{R}^{12}$ gauged supergravity}
\label{sec:4D_sugra}

\subsection{Maximal 4D supergravities: gaugings and scalar potential}

The bosonic sector of the maximal $\mathcal{N}=8$ supergravity multiplet in four dimensions consists of the spin-2 metric field $g_{\mu \nu}$, a set of $28$ spin-$1$ fields $\mathcal{A}_{\mu}{}^{\Lambda}$ with $\,\Lambda=1,\ldots 28$ (we will also introduce their magnetic duals $\tilde{\mathcal{A}}_{\mu \, \Lambda}$), and $70$ spin-$0$ fields serving as coordinates in a coset space ${\mathcal{M}_{\textrm{scal}}=\textrm{E}_{7(7)}/\textrm{SU}(8)}$. Electric and magnetic spin-$1$ fields can be arranged into
\begin{equation}
\label{elec/mag_vectors}
\mathcal{A}_{\mu}{}^{\mathbb{M}} = (\mathcal{A}_{\mu}{}^{\Lambda} , \tilde{\mathcal{A}}_{\mu \, \Lambda}) \ ,
\end{equation}
with $\mathbb{M}=1,\ldots,56$ being an index in the fundamental $\textbf{56}$ representation of $\textrm{E}_{7(7)}$: the global \textit{duality} symmetry underlying the four-dimensional \textit{ungauged} maximal supergravity. This symmetry acts non-linearly on the scalar fields (which are not charged under the $\textrm{U}(1)^{28}$ abelian gauge group spanned by the spin-$1$ fields) and, by virtue of supersymmetry, induces an $\textrm{Sp}(56)$ linear transformation on the spin-$1$ sector in (\ref{elec/mag_vectors}) that elevates the global $\textrm{E}_{7(7)}$ symmetry of the scalar sector to a symmetry of the field equations and Bianchi identities. In this manner, the index $\mathbb{M}$ is also identified with the $\textbf{56}$ of the $\textrm{Sp}(56)$ electromagnetic group of the theory.\footnote{We refer the reader to \cite{Trigiante:2016mnt} for a detailed review on this topic.}

We will be interested in the so-called \textit{gauged} maximal supergravities \cite{deWit:2007mt}. These are theories in which a non-abelian subgroup $\textrm{G}$ of the duality group $\textrm{E}_{7(7)}$ is promoted from global to local (gauge). In addition to the minimal couplings of the spin-$0$ fields to the spin-$1$ fields, consistency of the gauging procedure requires the introduction of a non-trivial scalar potential as well as scalar-dependent fermion mass terms in the Lagrangian. Moreover, for dyonic gaugings involving magnetic vector fields $\tilde{\mathcal{A}}_{\mu \, \Lambda}$, a topological term is also required together with a set of auxiliary two-form tensor fields which do not carry an independent dynamics (they are dual to spin-$0$ fields). All the modifications introduced by the gauging of $\textrm{G} \subset \textrm{E}_{7(7)}$  are encoded in an object called the \textit{embedding tensor} $\Theta_{\mathbb{M}}{}^{\alpha}$ where the index $\alpha=1,\ldots,133$ denotes the adjoint representation of $\textrm{E}_{7(7)}$. Consequently, the embedding tensor transforms as $\textbf{56} \times \textbf{133} = \textbf{912} + \ldots$ where the ellipsis stands for additional irreps that are projected out by the so-called linear or representation constraints of the theory \cite{deWit:2007mt}. In addition, consistency of the gauge structure requires an additional set of quadratic constraints of the form \cite{deWit:2007mt}
\begin{equation}
\label{QC_N8}
\Omega^{\mathbb{MN}} \, \Theta_{\mathbb{M}}{}^{\alpha}  \, \Theta_{\mathbb{N}}{}^{\beta}  = 0 \ ,
\end{equation}
where $\Omega^{\mathbb{MN}}$ is the skew-symmetric Sp(56) invariant matrix.

As discussed above, the embedding tensor $\Theta_{\mathbb{M}}{}^{\alpha} \in \textbf{912}$ of $\textrm{E}_{7(7)}$ fully determines the minimal couplings codified in the gauge connection as well as the scalar potential of a given maximal gauged supergravity. Covariant derivatives take the form
\begin{equation}
\label{cov_der}
D_{\mu} = \partial_{\mu} - \mathcal{A}_{\mu}{}^{\mathbb{M}} \, \Theta_{\mathbb{M}}{}^{\alpha}  \, t_{\alpha} \ ,
\end{equation}
with $t_{\alpha}$ being the generators of $\textrm{E}_{7(7)}$ in the appropriate representation. Note that (\ref{cov_der}) generically allows for magnetic vector fields to enter the gauge connection provided (\ref{QC_N8}) holds. The scalar potential of the theory is given by
\begin{equation}
\label{V(X)}
V(X,M) = \frac{1}{672} \,  {X_{\mathbb{MNP}}} \,  {X_{\mathbb{QRS}}} \, M^{\mathbb{MQ}} \big( M^{\mathbb{NR}}  \, M^{\mathbb{PS}} +   7 \,  \Omega^{\mathbb{RP}} \, \Omega^{\mathbb{NS}}  \big) \ ,
\end{equation}
in terms of the scalar-dependent matrix $M_\mathbb{MN}=(\mathcal{V} \, \mathcal{V}^{t})_\mathbb{MN}$ constructed from the coset representative $\mathcal{V}  \in \textrm{E}_{7(7)}/\textrm{SU}(8)$ and the $X$-tensor
\begin{equation}
\label{X-tensor}
X_{\mathbb{MN}}{}^{\mathbb{P}} = \Theta_{\mathbb{M}}{}^{\alpha}  \, [t_{\alpha}]_{\mathbb{N}}{}^{\mathbb{P}} \ ,
\end{equation}
which encodes the scalar charges entering the covariant derivative in (\ref{cov_der}). Recall that an index $\mathbb{M}$ is raised/lowered using the $\Omega^{\mathbb{MN}}$ and $\Omega_{\mathbb{MN}}$ matrices according to the North-West--South-East convention. These matrices satisfy $\Omega_{\mathbb{MN}} \, \Omega^{\mathbb{MP}} = \delta_{\mathbb{N}}^{\mathbb{P}}$.

It will also prove convenient to introduce a scalar-dependent $X$-tensor known as the \mbox{$T$-tensor }
\begin{equation}
T_{\mathbb{MNP}} = [ \mathcal{V}^{-1}] _{\mathbb{M}}{}^{\mathbb{Q}} \,    [ \mathcal{V}^{-1}] _{\mathbb{N}}{}^{\mathbb{R}} \,    [ \mathcal{V}^{-1}] _{\mathbb{P}}{}^{\mathbb{S}} \,   X_{\mathbb{QRS}} \ ,
\end{equation}
in terms of which the scalar potential in (\ref{V(X)}) is expressed as
\begin{equation}
\label{V(T)}
V(T) = \frac{1}{672} \,  {T_{\mathbb{MNP}}} \,  {T_{\mathbb{QRS}}} \, \delta^{\mathbb{MQ}} \big( \delta^{\mathbb{NR}}  \, \delta^{\mathbb{PS}} +   7 \,  \Omega^{\mathbb{RP}} \, \Omega^{\mathbb{NS}}  \big) \ .
\end{equation}
From (\ref{V(X)}) and (\ref{V(T)}) one finds that $V(X,M)=V(T)\,$. As for the $X$-tensor in (\ref{X-tensor}), the $T$-tensor can be written as
\begin{equation}
\label{Xi-tensor}
T_{\mathbb{MN}}{}^{\mathbb{P}} =\, \Xi_\mathbb{M}{}^{\alpha}  \, [t_{\alpha}]_{\mathbb{N}}{}^{\mathbb{P}}
\hspace{8mm} \text{ with } \hspace{8mm} 
\Xi_\mathbb{M}{}^{\alpha}  = \mathcal{V}^{-1} \star  \Theta_\mathbb{M}{}^{\alpha}  \ ,
\end{equation}
where the $\,\star\,$ operation denotes the action of a group element of $\textrm{E}_{7(7)}$ -- in this case the inverse coset representative $\mathcal{V}^{-1}$ -- both on the fundamental and the adjoint indices. In terms of the scalar-dependent $\Xi$-tensor in (\ref{Xi-tensor}) the scalar potential (\ref{V(T)}) can be rewritten as
\begin{equation}
\label{V(Xi)}
V(\Xi) 
% = \frac{1}{672}\,  \Xi_\mathbb{M}{}^{\alpha}  \,  \Xi_\mathbb{M}{}^{\beta}   \, \text{Tr}\left[t_\alpha (t_\beta^t + 7 t_\beta)\right] \ .
 = \frac{1}{672}\,  \Xi_\mathbb{M}{}^{\alpha}  \,  \Xi_\mathbb{M}{}^{\beta}   \, (\delta_{\alpha \beta} + 7 \, \mathcal{K}_{\alpha \beta}) \ ,
\end{equation}
where we made use of the normalisation condition $\,\textrm{Tr}(t_\alpha \, t_\beta^{t})=\delta_{\alpha \beta}\,$ and the definition of the Killing--Cartan matrix $\,\mathcal{K}_{\alpha \beta}=\textrm{Tr}(t_\alpha \, t_\beta)\,$. Note that the first contribution in the r.h.s of (\ref{V(Xi)}) is positive definite (sum of squares) while the second one has no definite sign due to the non-compactness of $\,\textrm{E}_{7(7)}\,$.

\subsection{$[\textrm{SO}(1,1) \times \textrm{SO}(6)] \ltimes \mathbb{R}^{12}$ dyonic gauging and S-folds}

Let us start by considering some group-theoretical branching rules of $\,\textrm{E}_{7(7)}\,$ irreps under its $\,\textrm{SL}(8) \subset \textrm{E}_{7(7)}\,$ maximal subgroup. Three representations are of particular interest: the fundamental $\,\mathbf{56}\,$ (index $\,\mathbb{M}$), the adjoint $\,\mathbf{133}\,$ (index $\,\alpha$), and the $\,\mathbf{912}\,$ (irrep of the embedding tensor). They decompose as
\begin{align}
\label{branching_56-SL(8)}
\mathbf{56}&\rightarrow \mathbf{28} + \mathbf{28}' \ , \\[1mm]
\label{branching_133-SL(8)}
\mathbf{133} &\rightarrow \mathbf{63} \oplus \mathbf{70} \ , \\[1mm]
\label{branching_912-SL(8)}
\mathbf{912}&\rightarrow\mathbf{36} \oplus \mathbf{420}\oplus \mathbf{36}' \oplus \mathbf{420}' \ .
\end{align} 
In terms of fundamental SL(8) indices $\,A=1,\ldots,8\,$ one has that 
\begin{equation}
\mathcal{A}^{\mathbb{M}} = ( \mathcal{A}^{[AB]}\, , \, \mathcal{A}_{[AB]})
\hspace{8mm} \textrm{ and } \hspace{8mm}
t_{\alpha} = ( t_{A}{}^{B} \, , \, t_{[ABCD]} ) \ ,
\end{equation}
with $\,t_{A}{}^{A}=0\,$ and where the square brackets denote antisymmetrisation. We have collected in the Appendix~\ref{app:E7-conventions} our conventions and various results regarding the $\,\textrm{E}_{7(7)}\,$ generators in the SL(8) basis.

In this note we will consider the dyonic gauging of $\,\textrm{G}=[\textrm{SO}(1,1) \times \textrm{SO}(6)] \ltimes \mathbb{R}^{12}$. This amounts to a choice of embedding tensor $\,\Theta_{\mathbb{M}}{}^{\alpha}\,$ of the form 
\begin{equation}
\label{Theta_tensor}
\Theta_{[AB]}{}^{C}{}_{D} = 4 \sqrt{3} \, \delta_{[A}^{C} \, \theta_{B]D}
\hspace{8mm} \textrm{ and } \hspace{8mm}
\Theta^{[AB]}{}^{C}{}_{D} = 4 \sqrt{3} \, \delta_{D}^{[A} \, \tilde{\theta}^{B]C} \ ,
\end{equation}
with
\begin{equation}
\label{etaMatrices}
\theta_{AB} = g \, \textrm{diag} (\, 0 \, , \, \mathbb{I}_{6}\, ,  \, 0 \,)
\hspace{8mm} \textrm{ and } \hspace{8mm}
\tilde{\theta}^{AB} = c \, \textrm{diag} (\, -1 \, , \, 0_{6}\, ,  \, 1 \,) \ ,
\end{equation}
so that only components belonging to the $\,\textbf{36}\,$ and the $\,\textbf{36}'\,$ are present in the decomposition (\ref{branching_912-SL(8)}). Note the presence of an electric parameter $\,g\,$ and a magnetic one $\,c\,$ rendering the gauging of dyonic type whenever $gc \neq 0$. The quadratic constraint in (\ref{QC_N8}) is automatically satisfied by the embedding tensor choice (\ref{Theta_tensor}) with (\ref{etaMatrices}).

The four-dimensional $\,[\textrm{SO}(1,1) \times \textrm{SO}(6)] \ltimes \mathbb{R}^{12}\,$ maximal gauged supergravity arises from the consistent reduction of type IIB supergravity on $\,\textrm{S}^{5}\,$ followed by a further $\,\textrm{S}^{1}\,$ reduction that incorporates an SL(2)$_{\textrm{IIB}}$ hyperbolic twist \cite{Inverso:2016eet}. This suggests to further decompose the $\,\textrm{E}_{7(7)}\,$ irreps in (\ref{branching_56-SL(8)})-(\ref{branching_912-SL(8)}) under the maximal subgroup $\,\textrm{SL}(6) \times \textrm{SL}(2) \times \textrm{SO}(1,1) \subset \textrm{SL}(8)\,$ in order to better analyse the gauging. The result is
\begin{align}
\label{branching_56-SL(6)xSL(2)}
	\mathbf{56} &\rightarrow \mathbf{28} \oplus \mathbf{28'}\\ 
	\nonumber&\rightarrow \left[(\mathbf{6},\,\mathbf{2})_{-1} \oplus (\mathbf{15},\,\mathbf{1})_{1} \oplus (\mathbf{1},\,\mathbf{1})_{-3}\right] \oplus\left[(\mathbf{6'},\,\mathbf{2})_{1} \oplus (\mathbf{15'},\,\mathbf{1})_{-1} \oplus (\mathbf{1},\,\mathbf{1})_{3}\right] \ , \\[2mm]
\label{branching_133-SL(6)xSL(2)}
  \mathbf{133} &\rightarrow \mathbf{63} \oplus \mathbf{70}\\
	\nonumber&\rightarrow \left[(\mathbf{35},\mathbf{1})_0 \oplus (\mathbf{6},\mathbf{2})_2\oplus(\mathbf{6'},\mathbf{2})_{-2} \oplus (\mathbf{1},\mathbf{3})_0 \oplus (\mathbf{1},\mathbf{1})_0\right] \\
\nonumber	&\phantom{\rightarrow}\hspace{5mm}\oplus \left[(\mathbf{15},\mathbf{1})_{-2} \oplus (\mathbf{20},\,\mathbf{2})_0 \oplus (\mathbf{15'},\mathbf{1})_2\right] \ , \\[2mm]
\label{branching_912-SL(6)xSL(2)}
\mathbf{912} &\rightarrow \mathbf{36} \oplus \mathbf{36'} \oplus \mathbf{420} \oplus \mathbf{420'}\\
	\nonumber&\rightarrow \left[(\mathbf{21},\mathbf{1})_1 \oplus (\mathbf{6},\mathbf{2})_{-1} \oplus (\mathbf{1},\mathbf{3})_{-3}\right] \oplus \left[(\mathbf{21'},\mathbf{1})_{-1} \oplus (\mathbf{6'},\mathbf{2})_{1} \oplus (\mathbf{1},\mathbf{3})_{3}\right] \\
	\nonumber&\phantom{\rightarrow} \oplus \left[(\mathbf{35},\mathbf{1})_{-3}\oplus (\mathbf{84},\mathbf{2})_{-1} \oplus (\mathbf{6},\mathbf{2})_{-1} \oplus (\mathbf{105},\mathbf{1})_{1} \oplus (\mathbf{15},\mathbf{3})_{1} \oplus (\mathbf{15},\mathbf{1})_{1} \oplus (\mathbf{20},\mathbf{2})_3 \right]\\
	\nonumber	&\phantom{\rightarrow} \oplus \left[(\mathbf{35},\mathbf{1})_{3}\oplus (\mathbf{84'},\mathbf{2})_{1} \oplus (\mathbf{6'},\mathbf{2})_{1} \oplus (\mathbf{105'},\mathbf{1})_{-1} \oplus (\mathbf{15'},\mathbf{3})_{-1} \oplus (\mathbf{15'},\mathbf{1})_{-1} \oplus (\mathbf{20},\mathbf{2})_{-3} \right] \ .
\end{align}
Under this decomposition the electric component of the embedding tensor corresponds to $\,{(\mathbf{21},\mathbf{1})_1 \subset \textbf{36}}\,$ whereas the magnetic one to $\,(\mathbf{1},\mathbf{3})_{3} \subset \textbf{36}'\,$. Similarly, the fundamental SL(8) index decomposes as $\,A = (i\,,\,a)\,$ with $\,{i=2,\ldots,7}\,$ being a fundamental SL(6) index and  $\,a=1,8\,$ a fundamental SL(2) index. As a consequence
\begin{equation}
[AB] = ([18] \,,\, [ai] \,,\, [ij]) \ ,
%\hspace{8mm} \textrm{ and } \hspace{8mm}
%[ABCD] = ([1ij8] \,,\, [aijk] \,,\, [ijkl]) \ ,
\end{equation}
and, from the embedding tensor components in (\ref{Theta_tensor})-(\ref{etaMatrices}), one gets
\begin{align}
\label{Theta_tensor_2}
\Theta_{\mathbb{M}}{}^{\alpha} \, t_{\alpha}
\,\,\,\, \rightarrow \,\,\,\,
&\Theta_{ij} = 2\sqrt{3}\,g \, ( t_{i}{}^{j} - t_{j}{}^{i} ) 
\hspace{3mm} , \hspace{3mm}
\Theta^{18} =2\sqrt{3}\, c \,\,  ( t_{1}{}^{8} + t_{8}{}^{1})\hspace{3mm}, \nonumber\\ 
&\Theta_{ai} = 2\sqrt{3}\,g \, t_{a}{}^{i}
\hspace{14mm} , \hspace{3mm}
\Theta^{ai} = 2\sqrt{3}\,c \,\,  t_{a}{}^{i} \ .
\end{align}
Then the compact SO(6) factor of the gauge group is gauged electrically, the non-compact SO(1,1) factor is gauged magnetically and the $\mathbb{R}^{12}$ translations are gauged dyonically provided $gc \neq 0$.

Various AdS$_{4}$ solutions have been found in this theory and subsequently uplifted to S-fold backgrounds of type IIB supergravity \cite{Inverso:2016eet,Guarino:2019oct,Guarino:2020gfe}. Being maximally symmetric vacua of the theory, vector and auxiliary tensor fields are set to zero resulting in a Lagrangian of Einstein-scalar type
\begin{equation}
\label{Lagrangian_N8}
\mathcal{L}  = \left(\frac{R}{2} - V\right) * 1 + \frac{1}{96} \text{Tr}\left(dM  \wedge * dM^{-1}\right) \ ,
\end{equation}
where $\,V\,$ is the scalar potential in (\ref{V(X)}) which is retained upon setting to zero vector and tensor fields. For the purpose of this note it will be convenient to use the Iwasawa or solvable parameterisation of the scalar coset representative 
\begin{equation}
\label{Iwasawa_param}
\mathcal{V} = \exp\left(\chi^{\hat{\texttt{e}}} \, t_{\hat{\texttt{e}}} \right)\,\cdot \, \exp\left(\phi^\texttt{h} \, t_\texttt{h}\right)\ ,
\end{equation}
where $\,t_\texttt{h}\,$ are the $\,7\,$ generators of the maximal non-compact torus $\,\mathbb{R}^7 \subset \textrm{E}_{7(7)}\,$ (Cartan subalgebra) and $\,t_{\hat{\texttt{e}}} \,$ are the $\,63\,$ positive roots of $\,\textrm{E}_{7(7)}\,$ (computed with respect to a choice of basis for the maximal torus). We will refer to the associated scalars as dilatons $\,\phi^\texttt{h}\,$ and axions $\,\chi^{\hat{\texttt{e}}}\,$. With the exception of the original $\,\mathcal{N}=4\,$ and $\,\textrm{SO}(4)\,$ symmetric S-fold in \cite{Gallerati:2014xra,Inverso:2016eet}, the rest of S-fold solutions presented in \cite{Guarino:2020gfe} within a $\,\mathbb{Z}_{2}^{3}$-invariant sector of the theory happened to contain moduli fields (axion-like fields $\,\chi^{\hat{\texttt{e}}}\,$) representing flat directions in the scalar potential. Understanding the general features of axion-like flat deformations of S-fold solutions will be the goal of this note.

\subsection{Axion-like flat deformations}

Let us consider a vacuum solution of the Lagrangian (\ref{Lagrangian_N8}), namely, an extremum of the scalar potential (\ref{V(X)}), and denote the associated coset representative at such a vacuum $\,\mathcal{V}_{0}\,$ and its vacuum energy $\,V_{0}\,$. Importantly, we will always assume some continuous residual symmetry group $\,\textrm{G}_{0} \subset \textrm{SO}(6) \subset \textrm{G}\,$ at the vacuum and parameterise an element of its algebra $\,\mathfrak{g}_{0}\,$ by an antisymmetric constant matrix $\,\chi^{ij}\,$. We will show that, starting from the G$_{0}$-invariant vacuum solution $\,\mathcal{V}_{0}\,$ of the theory, then the coset replacement
\begin{equation}
\label{Vchi_element}
\mathcal{V}_{0} \, \rightarrow \, \mathcal{V}_{\chi}  \mathcal{V}_{0}
\hspace{8mm} \textrm{ with } \hspace{8mm}
\mathcal{V}_{\chi} =  \exp \left(\tfrac{1}{2}\, \chi^{ij} \, t_{1ij8} \right) \in \textrm{E}_{7(7)} \ ,
\end{equation}
still describes a vacuum solution of the $\,\textrm{G}=[\textrm{SO}(1,1) \times \textrm{SO}(6)] \ltimes \mathbb{R}^{12}\,$ maximal gauged supergravity with the \textit{same} value of the cosmological constant $\,V_{0}\,$. The new vacuum solution $\,\mathcal{V}_{\chi} \mathcal{V}_{0}\,$ no longer belongs to the G$_{0}$-invariant sector of the theory, but nevertheless parameterises an entire family of $\chi$-dependent solutions with $\,\chi^{ij}\,$ corresponding to flat directions in the scalar potential. We will refer to the $\,\chi^{ij}\,$ parameters as axion-like flat deformations of the original S-fold solution $\,\mathcal{V}_{0}\,$.

\begin{table}[t]
\begin{center}
\renewcommand{\arraystretch}{1.5}
	\begin{tabular}{|c|c|c|c|c|}\hline
	$\otimes$&rep $\in \mathbf{912}$& $\theta$ & $\tilde{\theta}$ & $\delta \Theta$  \\
	$\mathbf{133}$   &  & $(\mathbf{21},\mathbf{1})_1$ & $(\mathbf{1},\mathbf{3})_3$ & $(\mathbf{35},\mathbf{1})_3$\\\hline
$t_{a}{}^{i}$	&$(\mathbf{6},\mathbf{2})_2$          & $\times$ & $\times$ & $\times$  \\\hline
	$t_{1ij8}$ &$(\mathbf{15'},\mathbf{1})_2$   & $(\mathbf{35},\mathbf{1})_3$  & $\times$ & $\times$ \\\hline
	$\mathfrak{sl}(2)$ &$(\mathbf{1},\mathbf{3})_0 $         & $\times$ & $(\mathbf{1},\mathbf{3})_3$ & $\times$  \\\hline
	$\mathfrak{sl}(6)$ &$(\mathbf{35},\mathbf{1})_0$         & $(\mathbf{21},\mathbf{1})_1 \oplus(\mathbf{15},\mathbf{1})_1$ & $\times$ & $(\mathbf{35},\mathbf{1})_3$  \\\hline
	$\mathfrak{so}(1,\,1)$ &$(\mathbf{1},\mathbf{1})_0$          & $(\mathbf{21},\mathbf{1})_1$ & $(\mathbf{1},\mathbf{3})_3$ & $(\mathbf{35},\mathbf{1})_3$ \\\hline
$t_{aijk}$	&$(\mathbf{20},\mathbf{2})_0 $        & $(\mathbf{84'},\mathbf{2})_1$ &$(\mathbf{20},\mathbf{2})_3$& $(\mathbf{20},\mathbf{2})_3$ \\\hline
$t_{i}{}^{a}$	&$(\mathbf{6'},\mathbf{2})_{-2}$ & $(\mathbf{6},\mathbf{2})_{-1}$ & $(\mathbf{6'},\mathbf{2})_{1}$ & $(\mathbf{84'},\mathbf{2})_1 \oplus (\mathbf{6'},\mathbf{2})_1$  \\\hline
$t_{ijkl}$	&$(\mathbf{15},\mathbf{1})_{-2}$      & $(\mathbf{105'},\mathbf{1})_{-1}$ & $(\mathbf{15},\mathbf{3})_1$ & $(\mathbf{105},\mathbf{1})_1 \oplus (\mathbf{21},\mathbf{1})_1 \oplus (\mathbf{15},\mathbf{1})_1$ \\\hline
\end{tabular}
\caption{Group-theoretical action of the $\mathbf{133}$ representation of $\,\textrm{E}_{7(7)}\,$ on the $\mathbf{912}$ using the $\,\textrm{SL}(6)\times \textrm{SL}(2) \times \textrm{SO}(1,1)\,$ basis. Only pieces belonging to the $\textbf{912}$ are displayed.}
\label{Table:133x912}
\end{center}
\end{table}

To prove the statement above we will take advantage of the $\textrm{E}_{7(7)}/\textrm{SU}(8)$ coset structure of the scalar manifold of maximal supergravity. As discussed in the introduction, the solution with non-zero axions $\,\mathcal{V}_{\chi}  \mathcal{V}_{0}\,$ in the $\textrm{G}$-gauged maximal supergravity can be mapped into an axion-vanishing solution in a \textit{different} theory with different gauge group $\,\widetilde{\textrm{G}} \neq \textrm{G}\,$ specified by an embedding tensor
\begin{equation}
\label{tildeTheta-tensor}
\widetilde{\Theta}_\mathbb{M}{}^{\alpha}  = \mathcal{V}_{\chi} \star  \Theta_\mathbb{M}{}^{\alpha} =  \Theta_\mathbb{M}{}^{\alpha} + (\delta\Theta)_\mathbb{M}{}^{\alpha} \ ,
\end{equation}
where the $\,\star\,$ denotes the action of the $\,\textrm{E}_{7(7)}\,$ element $\, \mathcal{V}_{\chi}\,$ in (\ref{Vchi_element}) on the embedding tensor $\, \Theta_\mathbb{M}{}^{\alpha} \in \textbf{912}\,$ of the original G-gauged theory. At the linear level, the element $\,\mathcal{V}_\chi \,$ is generated by $\,t_{1ij8} \in (\mathbf{15}',\,\mathbf{1})_{2}\,$. Then, an inspection of Table~\ref{Table:133x912} shows that, upon acting on the original embedding tensor components in (\ref{Theta_tensor}), the action of $\,\mathcal{V}_\chi\,$ only produces a single term $\,\delta\Theta \in (\mathbf{35},\,\mathbf{1})_3 \subset \textbf{420}'\,$ that originates from the electric piece $\,(\mathbf{21},\,\mathbf{1})_1 \subset \textbf{36}\,$ in the $\mathbf{912}$ decomposition (\ref{branching_912-SL(6)xSL(2)}).\footnote{This implies that our argument holds in the limit where $\,c\rightarrow 0\,$ as well as for the magnetic gauging of an $\,\textrm{SO}(2)\,$ factor instead of $\,\textrm{SO}(1,1)\,$.} Equivalently, only a linear term appears since the $\,\textrm{SO}(1,1)\,$ grading in the decomposition (\ref{branching_912-SL(6)xSL(2)}) solely allows for charges $\,-3\,,\,-1,\,1\,$ or $\,+3\,$, and not $\,+5\,$ or higher which would be the ones produced beyond the linear level.  As a result, $\,\delta \Theta\,$ is invariant under the action of $\,\mathcal{V}_{\chi}\,$. An explicit computation of $\,\delta\Theta\,$ in (\ref{tildeTheta-tensor}) yields
\begin{equation}
\label{deltaTheta-tensor}
(\delta\Theta)_{\mathbb{M}}{}^{\alpha} \, t_{\alpha}
\,\,\,\, \rightarrow \,\,\,\, 
\delta \Theta_{ij} = 2 \, \chi_{[i|}{}^{k} \,\, t_{{1k|j]8}}
\hspace{3mm} , \hspace{3mm}
\delta \Theta^{a i} =  -\, \epsilon^{ab}  \, \chi_{j}{}^{i} \, t_{b}{}^{j}
\hspace{3mm} , \hspace{3mm}
\delta \Theta^{18} =  \,  \chi_j{}^i  \, t_{i}{}^{j} \ ,
\end{equation}
where $\,\chi_{i}{}^{j}= \, \theta_{ik} \, \chi^{kj}\,$ with $\,\theta_{ij}= g \, \delta_{ij}\,$ in (\ref{etaMatrices}). This comes from the fact that the only terms in the product $\,\mathbf{56}\otimes\mathbf{133}\,$ contributing to the $\,(\mathbf{35}, \mathbf{1})_3\,$ are precisely the $\,(\mathbf{15}, \mathbf{1})_1 \otimes (\mathbf{15'},\mathbf{1})_2\,$, the $\,(\mathbf{6'},\mathbf{2})_1\otimes (\mathbf{6},\mathbf{2})_2\,$ and the $\,(\mathbf{1},\mathbf{1})_3 \otimes (\mathbf{35},\mathbf{1})_0\,$. It is also worth noticing that $\,\delta\Theta\,$ in (\ref{deltaTheta-tensor}) verifies the quadratic constraints (\ref{QC_N8}) so it defines a consistent gauging by itself.

Using $\,\widetilde{\Theta}_\mathbb{M}{}^{\alpha}\,$ in (\ref{tildeTheta-tensor}) with $\,\Theta_\mathbb{M}{}^{\alpha}\,$ and $\,(\delta\Theta)_\mathbb{M}{}^{\alpha} \,$ given in (\ref{Theta_tensor_2}) and (\ref{deltaTheta-tensor}) respectively, it can be proved that
\begin{equation}
\label{V=V=V}
V(\Theta ,\mathcal{V}_{\chi} \mathcal{V}_{\textrm{G$^\chi_{0}$-inv}}) = V(\widetilde{\Theta} ,\mathcal{V}_{\textrm{G$^\chi_{0}$-inv}}) = V(\Theta , \mathcal{V}_{\textrm{G$^\chi_{0}$-inv}}) \ ,
\end{equation} 
where $\,\textrm{G}_0^\chi\,$ denotes a Cartan subgroup of $\,\mathrm{G}_0\,$ commuting with $\,\chi_{i}{}^{j}\,$ and $\,\mathcal{V}_{\textrm{G$^\chi_{0}$-inv}}\,$ denotes the scalar-dependent coset representative of the $\,\textrm{G}^\chi_{0}$-invariant sector of maximal supergravity (which contains the G$_{0}$-invariant vacuum $\,\mathcal{V}_{0}\,$). This implies that there always exist flat directions of the scalar potential at the vacuum $\,\mathcal{V}_{0}\,$ parameterised by the axions $\,\chi^{ij}\,$ that are \textit{not} captured by the G$_{0}$-invariant sector of the theory. The first equality in (\ref{V=V=V}) is a direct consequence of the formally $\textrm{E}_{7(7)}$-covariant formulation of the maximal 4D gauged supergravities provided by the embedding tensor formalism. A detailed proof of the second equality in (\ref{V=V=V}) is presented in the Appendix~\ref{app:Proof}.

\subsection{Deforming the $\,\mathcal{N}=4\,$ and $\,\textrm{SO}(4)\,$ symmetric S-fold}

A direct consequence of (\ref{V=V=V}) is the existence of two axion-like flat deformations\footnote{The index $\,\alpha\,$ in this section should not be confused with the $\,\textrm{E}_{7(7)}\,$ adjoint index in the previous sections.} $\,\chi_{\alpha}\,$ (${\alpha=1,2}$) of the original $\,\mathcal{N}=4\,$ S-fold with $\,\textrm{G}_{0}=\textrm{SO}(4)\,$ symmetry, which control the pattern of symmetry breaking down to its $\,\textrm{G}^{\chi}_{0}=\textrm{U}(1)^{2}\,$ Cartan subgroup. These flat deformations lie outside the $\,\mathbb{Z}_{2}^{3}$-invariant sector of the theory investigated in \cite{Guarino:2020gfe} and, adopting the conventions therein, they specify a matrix $\,\chi^{ij}\,$ of the form
\begin{equation} 
\label{chi_matrix_N4}
\chi^{ij} = 12 \, \sqrt{2} \,  \left(
\begin{array}{cccccc}
0 & 0 & 0 & 0 & 0 & 0 \\
0 & 0 & 0 & 0 & 0 & 0 \\
0 & 0 & 0 & 0 & \chi_{1} & \chi_{2} \\
0 & 0 & 0 & 0 & \chi_{2} & \chi_{1} \\
0 & 0 & -\chi_{1}  & -\chi_{2} & 0 & 0 \\
0 & 0 & -\chi_{2} & -\chi_{1}  & 0 & 0
\end{array}
\right) \in \mathfrak{so}(2)^2  \subset \mathfrak{so}(4) \ .
\end{equation}
An explicit computation of the full scalar and vector mass spectra at the corresponding AdS$_{4}$ vacua yields the following results. The normalised spectrum (masses and multiplicities) of scalar fields is given by
\begin{equation}
\label{spectrum_scalars}
\begin{array}{lll}
m^2 L^2 &=& 10 \,\, (\times 1) \,\,\, , \,\,\, 4 \,\, (\times 2) \,\,\, , \,\,\, -2 \,\, (\times 3) \,\,\, , \,\,\, 0 \,\, (\times 32) \,\,\, , \,\,\, \chi_{\alpha}^2 \,\, (\times 2) \ , \\[2mm]
& & (\chi_{1} \pm \chi_{2})^2 \,\, (\times 2)  \,\,\, , \,\,\,   \frac{1}{4} \, (\chi_{1} \pm \chi_{2})^2 \,\, (\times 4) \,\,\, , \,\,\, 
1+ \chi_{\alpha}^2 \pm \sqrt{9 + 4 \,  \chi_{\alpha}^2} \,\, (\times 2) \ , \\[3mm]
& & 1+ \frac{1}{4} \, (\chi_{1}+\chi_{2})^2 \pm \sqrt{9 + (\chi_{1}+\chi_{2})^2} \,\, (\times 2)  \ , \\[3mm]
& & 1+ \frac{1}{4} \, (\chi_{1}-\chi_{2})^2 \pm \sqrt{9 + (\chi_{1}-\chi_{2})^2} \,\, (\times 2)  \ ,
\end{array}
\end{equation}
in terms of the AdS$_{4}$ radius  $\,L^2=-3/V_{0} = g^{-2} c\,$. The normalised spectrum (masses and multiplicities) of vector fields reads
\begin{equation}
\label{spectrum_vectors}
\begin{array}{lll}
m^2 L^2 &=& 0 \,\, (\times 2) \,\,\, , \,\,\,   2 \,\, (\times 3) \,\,\, , \,\,\,  6 \,\, (\times 3) \,\,\, , \,\,\,  2 + \chi_{\alpha}^2 \,\, (\times 2) \,\,\, , \,\,\, 2 + \frac{1}{4}(\chi_{1} \pm \chi_{2})^2 \,\, (\times 4)  \ , \\[3mm]
& & 3 + \frac{1}{4} \, (\chi_{1}+\chi_{2})^2 \pm \sqrt{9 + (\chi_{1}+\chi_{2})^2} \,\, (\times 2) \ , \\[3mm]
& & 3 + \frac{1}{4} \, (\chi_{1}-\chi_{2})^2 \pm \sqrt{9 + (\chi_{1}-\chi_{2})^2} \,\, (\times 2) \ ,
\end{array}
\end{equation}
and contains two massless vectors at generic values of $\,\chi_{\alpha}\,$. Lastly, the computation of the eight normalised gravitino masses yields\footnote{In our conventions a massless gravitino associated with a preserved supersymmetry has $\,m^2 L^2=1\,$.}
\begin{equation}
\label{spectrum_gravitini}
\begin{array}{lll}
m^2 L^2 &=& \frac{5}{2} +\frac{1}{4} \chi_{\alpha}^2  \pm  \frac{1}{2} \sqrt{9+\chi_{\alpha}^{2}}    \,\, (\times 2)   \ .
\end{array}
\end{equation}

By inspection of (\ref{spectrum_scalars})-(\ref{spectrum_gravitini}) we identify four different classes of flat deformations of the $\,\mathcal{N} = 4\,$ and $\,\textrm{SO}(4)\,$ symmetric S-fold  (see Figure~\ref{fig:diagram}):
\begin{itemize}

\item[$\circ$] At generic values of $\,\chi_{1,2}\,$ one finds non-supersymmetric S-folds with a $\,\textrm{U}(1)^2\,$ symmetry which is interpreted as a flavour symmetry in the dual S-fold CFT's.

\item[$\circ$] Setting $\,\chi_{2}=0\,$ (equivalently $\,\chi_{1}=0$) produces a one-parameter family of $\,\mathcal{N}=2\,$ supersymmetric S-folds with $\,\textrm{U}(1) \times \textrm{U}(1) \,$ symmetry. These holographically describe a subclass (see footnote \ref{footnote_N2_manifold}) of the $\,\mathcal{N}=2\,$ S-fold CFT's with a $\,\textrm{U}(1)\,$ flavour symmetry investigated in \cite{Bobev:2021yya}. 

\item[$\circ$] Setting $\,\chi_{1} = \pm \chi_{2}\,$ gives rise to a one-parameter family of non-supersymmetric S-folds with $\,\textrm{SU}(2) \times \textrm{U}(1)\,$ symmetry. This implies a flavour symmetry enhancement of the form $\,\textrm{U}(1) \times \textrm{U}(1) \rightarrow \textrm{SU}(2) \times \textrm{U}(1)\,$ in the dual S-fold CFT's.

\item[$\circ$] Setting $\,\chi_{1} = \chi_{2} = 0\,$ gives back the original (undeformed) $\,\mathcal{N}=4\,$ supersymmetric S-folds with $\,\textrm{SO}(4)\,$ symmetry. 

\end{itemize}
Note that $\,\chi_{1}\,$ and $\,\chi_{2}\,$ enter (\ref{spectrum_scalars})-(\ref{spectrum_gravitini}) symmetrically, as expected, and that the scalar mass spectrum in (\ref{spectrum_scalars}) does not display any instability associated with a normalised mass mode violating the Breitenlohner--Freedman (BF) bound $\,{m^2 L^2 \ge -9/4}\,$ for perturbative stability in AdS$_{4}$ \cite{Breitenlohner:1982bm}. Therefore, all the non-supersymmetric S-folds discussed above turn out to be perturbatively stable at the lower-dimensional supergravity level.

Carrying out a study of their higher-dimensional stability by computing the associated Kaluza--Klein (KK) spectra, as done in \cite{Giambrone:2021zvp} (following \cite{Malek:2019eaz}) for the $\,\mathcal{N}=2\,$ S-folds of \cite{Guarino:2020gfe} and the $\,\mathcal{N}=4\,$ S-fold of \cite{Inverso:2016eet}, would help in establishing the perturbative (in)stability of the non-supersymmetric S-folds.\footnote{See \cite{Malek:2020mlk,Guarino:2020flh} for the KK spectrometry of M-theory and massive type IIA non-supersymmetric AdS$_{4}$ vacua.} The reason to perform such a KK spectroscopy study is two-fold. Firstly, to search for (super) symmetry enhancements when considering modes up in the KK tower. Secondly, to give support or pose a new challenge to the AdS Swampland Conjecture~\cite{Ooguri:2016pdq}.

\section{5D origin and CSS gaugings}
\label{sec:5D_CSS}

The results in the previous section followed from the very specific flat deformation introduced by the axions $\,\chi^{ij}\,$ which, as already emphasised, can be alternatively understood in terms of the $\,\delta\Theta\,$ deformation tensor in (\ref{deltaTheta-tensor}). The attentive reader might have recognised in (\ref{deltaTheta-tensor}) a structure similar to a CSS gauging \cite{Cremmer:1979uq}. This type of gaugings appear upon compactification of 5D supergravity down to 4D. We will now elaborate more on this point.

Let us start by recalling that there is a formally $\textrm{E}_{6(6)}$-covariant formulation of the maximal 5D gauged supergravities provided by the embedding tensor formalism \cite{deWit:2004nw}. The bosonic sector of the maximal $\,\mathcal{N}=8\,$ supergravity multiplet in five dimensions consists of the metric field $\,g_{\mu \nu}\,$, a set of $\,\textbf{27}'\,$ vector fields and $\,\textbf{27}\,$ two-form tensor fields, and $\,42\,$ scalar fields serving as coordinates in a coset space $\,{\mathcal{M}_{\textrm{scal}}=\textrm{E}_{6(6)}/\textrm{USp}(8)}$. The embedding tensor is subject to a set of linear or representation constraints restricting it to the $\,\textbf{351}\,$ representation. In addition it must also obey a set of quadratic constraints in order to specify a consistent gauging of the theory. To establish a 4D $\leftrightarrow$ 5D connection we will perform a group-theoretical decomposition of the $\,\textrm{E}_{7(7)}\,$ representations in (\ref{branching_56-SL(6)xSL(2)}), (\ref{branching_133-SL(6)xSL(2)}) and (\ref{branching_912-SL(6)xSL(2)}) under the maximal subgroup $\,\textrm{E}_{6(6)} \times \textrm{SO}(1,1) \subset \textrm{E}_{7(7)}\,$. This yields
\begin{align}
\label{branching_56-E6}
\mathbf{56} &\rightarrow \mathbf{1}_{-3} \oplus \mathbf{27}'_{-1} \oplus \mathbf{27}_{+1} \oplus \mathbf{1}_{+3}  \ , \\[1mm]
\label{branching_133-E6}
\mathbf{133} &\rightarrow \mathbf{78}_{0} \oplus \mathbf{1}_0 \oplus \mathbf{27}_{-2} \oplus \mathbf{27}'_{+2} \ , \\[1mm]
\label{branching_912-E6}
\mathbf{912} &\rightarrow \mathbf{78}_{-3} \oplus\mathbf{27}'_{-1} \oplus\mathbf{351}'_{-1} \oplus \mathbf{351}_{+1} \oplus\mathbf{27}_{+1} \oplus\mathbf{78}_{+3} \ .
\end{align}
In (\ref{branching_912-E6}) we observe the embedding tensor $\,\mathbf{351}_{+1}\,$ of the 5D theory descending from the embedding tensor $\,\textbf{912}\,$ of the 4D theory. However, in order to understand the embedding tensor $\,\tilde{\Theta}\,$ in (\ref{tildeTheta-tensor}), we will further need the $\,\mathbf{78}_{+3} \,$ in (\ref{branching_912-E6}). This becomes clear when looking at the group-theoretical decompositions of the said representations under $\,\textrm{E}_{6(6)} \times \textrm{SO}(1,1) \rightarrow \textrm{SL}(6) \times \textrm{SL}(2) \times \textrm{SO}(1,1)$, namely,
\begin{align}
\label{branching_351-SL6xSL2}
	\mathbf{351}_{+1} & \rightarrow (\mathbf{21},\,\mathbf{1})_{+1} \oplus (\mathbf{84'},\,\mathbf{2})_{+1}\oplus(\mathbf{105},\,\mathbf{1})_{+1} \oplus (\mathbf{15},\,\mathbf{3})_{+1} \oplus (\mathbf{6'},\,\mathbf{2})_{+1} \ , \\[1mm]
\label{branching_78-SL6xSL2}
	\mathbf{78}_{+3} &\rightarrow (\mathbf{35},\,\mathbf{1})_{+3} \oplus (\mathbf{20},\,\mathbf{2})_{+3} \oplus (\mathbf{1},\,\mathbf{3})_{+3} \ .
	%\label{branching_27-SL6xSL2}
	%\mathbf{27} &\rightarrow (\mathbf{6}',\,\mathbf{2}) \oplus  (\mathbf{15},\,\mathbf{1}) \ .
\end{align}
Importantly, while the $\,\mathbf{351}_{+1}\,$ captures the electric $\,(\mathbf{21},\mathbf{1})_{+1}\,$ piece in the 4D embedding tensor $\widetilde{\Theta}$ induced by $\,g\,$, it does not capture the magnetic $\,(\mathbf{1},\mathbf{3})_{+3}\,$ and $\,(\mathbf{35},\mathbf{1})_{+3}\,$ pieces induced by $\,c\,$ and $\,\chi_{i}{}^{j}\,$. These two pieces are instead contained in the $\,\mathbf{78}_{+3}\,$, as it is clear from the SO(1,1) charge. A direct consequence is that the 4D gauging $\widetilde{\Theta}$ involving the $\, (\mathbf{1},\,\mathbf{3})_{+3}  \subset \mathbf{78}_{+3} \,$ ($c$-terms) and $\, (\mathbf{35},\,\mathbf{1})_{+3}  \subset \mathbf{78}_{+3} \,$ ($\chi$-terms) cannot be directly uplifted to an embedding tensor deformation in 5D. These terms are instead generated dynamically by introducing an explicit dependence on the S$^{1}$ coordinate (in the form of a duality twist \cite{Dabholkar:2002sy}) in the reduction process from 5D to 4D.

A general duality twist takes the form \cite{Cremmer:1979uq,Scherk:1979zr}
\begin{equation}
\label{duality_twist}
\phi(x^\mu,\,\eta) = e^{M \,\eta} \star \phi(x^\mu) \ ,
\end{equation}
where $\,\eta\,$ is the coordinate along the S$^{1}$,  $\,M \in \mathfrak{e}_{6(6)}\,$ and $\,\phi\,$ is any field in the 5D theory. This type of duality twists has been studied in the context of 5D \textit{ungauged} supergravity \cite{Andrianopoli:2002mf}. Within this context, the dependence on the $\,\eta\,$ coordinate factorises out in the reduction process as a consequence of $\,M\,$ being chosen in the \emph{global} duality group $\,\textrm{E}_{6(6)}\,$ of the 5D theory. Moreover, choosing $\,M\,$ in the maximal compact subalgebra $\,\mathfrak{usp}(8) \subset \mathfrak{e}_{6(6)}\,$ makes the scalar potential vanish identically in the reduced 4D theory. Our scenario, however, differs from the one just discussed: the theory to begin with is the 5D SO(6)-gauged supergravity. The global duality group $\,\textrm{E}_{6(6)}$ is broken to a \emph{local} SO(6) and a \emph{global} SL(2), and the axion-like flat deformation $M = \chi_i{}^j \in \mathfrak{g}_{0} \subset \mathfrak{so}(6)\,$ leaves invariant the embedding tensor of the SO(6) gauging. Then, in our case, the flat deformations $\,\chi_{i}{}^{j}\,$ are expected to describe trivial twists leaving the putative 5D backgrounds locally invariant.

The simultaneous study of axion-like $\,\chi_{i}{}^{j}\,$ and electromagnetic $\,c\,$ deformations requires to investigate a general twist in the $\,\mathbf{78}_{+3}\,$.  From a 4D perspective, and by virtue of (\ref{branching_78-SL6xSL2}), the most general contraction $\,(\delta\Theta)_{\mathbb{M}}{}^{\alpha} \, t_{\alpha}\, \in \mathbf{78}_{+3}$ takes the form
\begin{equation}
\begin{array}{lll}
\label{general_tildeTheta}
\delta\Theta_{ij} &=&   \chi_{i}{}^k \, t_{1kj8}  - \chi_{j}{}^k \, t_{1ki8}  + \chi^{almn} \, t_a{}^k \,  \epsilon_{kijlmn}\ , \\[2mm]
\delta\Theta^{ai} &=& - \epsilon^{ab} \, \chi_j{}^i  \, t_b{}^j  - \epsilon^{ab} \, \chi_b{}^c \, t_c{}^i - 3 \, \epsilon^{ab} \,  \chi^{cijk} \,  t_{bjkc}   \ , \\[2mm]
\delta\Theta^{18} &=& \chi_i{}^j \, t_j{}^i + \chi_a{}^b \, t_b{}^a  + \chi^{aijk} \, t_{aijk}  \ ,
\end{array}
\end{equation}
in terms of three tensors $\,\chi_{i}{}^{j} \in (\textbf{35},\textbf{1})_{+3}\,$, $\,\chi_{a}{}^{b}\in (\textbf{1},\textbf{3})_{+3}\,$ and $\,\chi^{aijk}=\chi^{a[ijk]} \in  (\mathbf{20},\,\mathbf{2})_{+3} \,$. The embedding tensor $\,\delta\Theta\,$ in (\ref{general_tildeTheta}) satisfies the quadratic constraint (\ref{QC_N8}) and therefore defines a consistent gauging of the maximal supergravity in 4D. Since $\,\delta\Theta \in \mathbf{78}_{+3} \,$ carries the highest SO(1,1) charge in the decomposition (\ref{branching_912-E6}) and, when using the solvable parameterisation of $\textrm{E}_{7(7)}/\textrm{SU}(8)$, the coset representative $\,\mathcal{V}\,$ in (\ref{Iwasawa_param}) solely involves $\,\textrm{E}_{7(7)}\,$ generators with non-negative SO(1,1) charge (Cartan generators and positive roots), it follows that $\,\delta\Xi \in \mathbf{78}_{+3} \,$. Then, as for $\,\delta\Theta\,$, the scalar-dependent $\,\delta\Xi \,$ tensor can be expressed in terms of three scalar-dependent tensors $\,\delta\Xi_{i}{}^{j} \in (\textbf{35},\textbf{1})_{+3}\,$, $\,{\delta\Xi'_{a}{}^{b}\in (\textbf{1},\textbf{3})_{+3}}\,$ and $\,\delta\Xi^{aijk}=\delta\Xi^{a[ijk]} \in  (\mathbf{20},\,\mathbf{2})_{+3} \,$.

An explicit computation of the scalar potential (\ref{V(Xi)}) gives
\begin{equation}
\label{V_78}
V = \frac{1}{192} \Bigg\{ \text{Tr}\Big[ \left(\delta\Xi+(\delta\Xi)^t \right)^2 \Big] +   \text{Tr}\Big[ \left(\delta\Xi'+(\delta\Xi')^t \right)^2 \Big]  +  3 \sum\limits_{a\,i\,j\,k} \big( \delta\Xi^{aijk} - \tfrac{1}{3!} \, \epsilon_{ab} \, \epsilon_{ijklmn}  \,\delta\Xi^{blmn}\big)^2 \Bigg\} \ ,
\end{equation}
which, as expected, is positive definite. From the potential in (\ref{V_78}) we see that having an $\,\mathfrak{so}(6)$-valued $\,\delta\Xi_{i}{}^{j}\,$ implies $\,\delta\Xi+(\delta\Xi)^t=0\,$ and thus a flat CSS deformation with $V=0$. This is only possible if $\,\chi_{i}{}^{j} = - \chi_{j}{}^{i}\,$. On the contrary, having an $\,\mathfrak{so}(1,1)$-valued $\,\delta\Xi'_{a}{}^{b}\,$ yields $\,\delta\Xi' + (\delta\Xi')^t \neq 0\,$ and therefore $\,V \neq 0\,$. This explains why the axions $\,\chi^{ij}\,$ specifying a compact $\,\mathfrak{so}(6)\,$ twist produce flat deformations whereas the magnetic parameter $\,c\,$ specifying a non-compact $\,\mathfrak{so}(1,1)\,$ twist does not. In addition, there is the contribution to the scalar potential (\ref{V_78}) coming from $\,\delta\Xi^{aijk} \in  (\mathbf{20},\,\mathbf{2})_{+3} \,$. The $\,20\,$ linear combinations satisfying $\,\delta\Xi^{aijk} - \tfrac{1}{3!} \, \epsilon_{ab} \, \epsilon_{ijklmn}  \,\delta\Xi^{blmn}=0\,$ belong to $\,\mathfrak{usp}(8) \in \mathfrak{e}_{6(6)}\,$ and, in combination with $\,\delta\Xi_{i}{}^{j} \in \mathfrak{so}(6)\,$ and $\,\delta\Xi'_{a}{}^{b} \in \mathfrak{so}(2)\,$, they specify the most general flat CSS gauging yielding $\,V=0\,$. It would be interesting to further investigate the parameters $\,\delta\Xi^{aijk}\,$ and their potential to generate new type IIB backgrounds.

To conclude, group-theoretical arguments put forward in \cite{Guarino:2021kyp} suggest that axion-like deformations should be related to one-form deformations of $\,\mathcal{N}=4\,$ SYM on $\,{\mathbb{R}^{1,2}\times \textrm{S}^1}\,$ \cite{Maxfield:2016lok}. Such one-forms are often discarded within the context of Janus solutions by a gauge-fixing argument without much regard for large gauge transformations. Working out the explicit 5D oxidation of the AdS$_{4}$ vacua with $\,\chi^{ij} \neq 0\,$ would be the next step towards testing these ideas. In addition, it would also be interesting to further investigate the interplay between embedding tensor deformations $\,(\textbf{351}_{+1})\,$ and $\,\mathfrak{e}_{6(6)}\,$ duality twists $\,(\textbf{78}_{+3})\,$ in $\,\textrm{S}^{1}\,$ compactifications of other 5D maximal supergravities. Also to understand the physical meaning (if any) of the rest of representations appearing in the group-theoretical decomposition of the $\,\textbf{912}\,$ in (\ref{branching_912-E6}). We leave these and other related questions for future work.

\section*{Acknowledgements}

We thank Nikolay Bobev, Emanuel Malek and Jesse van Muiden for discussions. The work of AG is supported by the Spanish government grant PGC2018-096894-B-100. The research of CS is supported by IISN-Belgium (convention 4.4503.15). CS is a Research Fellow of the F.R.S.-FNRS (Belgium).

\appendix

\section{Conventions on $\textrm{E}_{7(7)}$ generators}
\label{app:E7-conventions}

In the SL(8) basis the adjoint representation of $\,\textrm{E}_{7(7)}\,$ splits into $\,\textbf{133} \rightarrow \textbf{63} \oplus \textbf{70}\,$  under $\,\textrm{SL}(8)\subset \textrm{E}_{7(7)}\,$. This implies a splitting of generators of the form $\,t_{\alpha} \rightarrow t_{A}{}^{B} \oplus t_{ABCD}\,$ with $\,t_{A}{}^{A}=0\,$ and $\,t_{ABCD} = t_{[ABCD]}\,$. The fundamental representation of $\,\textrm{E}_{7(7)}\,$ branches as $\,\textbf{56} \rightarrow \textbf{28} \oplus \textbf{28}'\,$ so the fundamental $\,\textrm{E}_{7(7)}\,$ index splits as $\,_{\mathbb{M}} \rightarrow _{[AB]} \oplus ^{[AB]}\,$. Then, the $\,\textbf{63}\,$ generators of SL(8) correspond with $\,\textrm{E}_{7(7)}\,$ generators of the form
\begin{equation}
\label{E7_gen_63}
[t_A{}^B]_\mathbb{M}{}^\mathbb{N} = \frac{1}{\sqrt{12}}
\begin{pmatrix}2 \, \delta^{[E}_{[C} \, [t_A{}^B]_{D]}{}^{F]} & 0 \\ 
0 & - 2 \,\delta^{[C}_{[E} \,\, [t_A{}^B]_{F]}{}^{D]}  \end{pmatrix} 
\hspace{3mm} \text{   with   } \hspace{3mm}
[t_A{}^B]_C{}^D = 4 \, \delta_{A}^C \, \delta_D^B - \frac{1}{2} \,  \delta_A^B \, \delta_C^D \ ,
\end{equation}
whereas the remaining $\,\textbf{70}\,$ generators extending SL(8) to E$_{7(7)}$ take the form
\begin{equation}
\label{E7_gen_70}
[ t_{ABCD} ]_\mathbb{M}{}^\mathbb{N} = \sqrt{12} \begin{pmatrix}0 & \epsilon_{ABCDEFGH} \\ 
4! \,\delta^{EFGH}_{ABCD} & 0\end{pmatrix} \ .
\end{equation}
They are normalized such that $\,\text{Tr}(t_\alpha t_\beta{}^t) = \delta_{\alpha\beta}\,$. The Killing-Cartan matrix is then given by 
\begin{equation}
\label{E7_KC}
\mathcal{K}_{\alpha \beta} =\text{Tr}(t_\alpha t_\beta) = 
\left\lbrace 
\begin{array}{l}
1 \hspace{5mm} \textrm{ if } \hspace{5mm} \beta = \alpha^t \\[2mm]
0 \hspace{5mm} \textrm{ otherwise} 
\end{array} 
\right.  \ ,
\end{equation}
where by $\,\alpha^t\,$ we refer to the generator $\,t_{\alpha^t} \equiv (t_\alpha)^t\,$. With the generators in (\ref{E7_gen_63}) and (\ref{E7_gen_70}) one has that
\begin{equation}
(t_A{}^B)^t = t_B{}^A
\hspace{8mm} \textrm{ and } \hspace{8mm} 
(t_{ABCD})^t = \frac{1}{4!} \, \epsilon^{ABCDEFGH} \, t_{EFGH} \ .
\end{equation}
Note that if $\,t_{\alpha}\,$ is a positive root of the $\,\mathfrak{e}_{7(7)}\,$ algebra then $\,t_{\alpha^{t}}\,$ is the corresponding negative root. As a result, the second term in the scalar potential (\ref{V(Xi)}) only receives contributions from pairs $\,(\Xi_\mathbb{M}{}^{\alpha}  ,  \Xi_\mathbb{M}{}^{\alpha^{t}})\,$ with opposite SO(1,1) charges in the decomposition (\ref{branching_912-SL(6)xSL(2)}).

\section{Proof of (\ref{V=V=V})}
\label{app:Proof}

In this appendix we present a proof of the relation (\ref{V=V=V}), namely,
\begin{equation}
\label{V=V=V_appendix}
V(\Theta ,\mathcal{V}_{\chi} \mathcal{V}_{\textrm{G$^\chi_{0}$-inv}}) = V(\widetilde{\Theta} ,\mathcal{V}_{\textrm{G$^\chi_{0}$-inv}}) = V(\Theta , \mathcal{V}_{\textrm{G$^\chi_{0}$-inv}}) \ ,
\end{equation} 
where $\,\widetilde{\Theta}_\mathbb{M}{}^{\alpha}\,$ is given in (\ref{tildeTheta-tensor}) in terms of the tensors $\,\Theta_\mathbb{M}{}^{\alpha}\,$ and $\,(\delta\Theta)_\mathbb{M}{}^{\alpha} \,$ in (\ref{Theta_tensor_2}) and (\ref{deltaTheta-tensor}), respectively, and where $\mathcal{V}_{\textrm{G$^\chi_{0}$-inv}}$ denotes the coset representative for the G$^\chi_{0}$-invariant sector of maximal supergravity with $\,\textrm{G}^\chi_{0} \subset \textrm{G}_{0} \subset \textrm{SO}(6)\,$. As already emphasised in the main text, the first equality in (\ref{V=V=V_appendix}) follows from the  $\textrm{E}_{7(7)}$-covariant formulation of the maximal 4D gauged supergravities provided by the embedding tensor formalism. Before proving the second equality in (\ref{V=V=V_appendix}), let us note that, since $\,\widetilde{\Theta} = \Theta + \delta\Theta\,$, one has
\begin{equation}
\label{Vtilde_terms}
V(\widetilde{\Theta} ,\mathcal{V}_{\textrm{G$^\chi_{0}$-inv}}) 
= V(\Theta ,\mathcal{V}_{\textrm{G$^\chi_{0}$-inv}}) \,+\,  2 \, B(\Theta , \delta\Theta  \, ; \, \mathcal{V}_{\textrm{G$^\chi_{0}$-inv}})  \,+\,  V(\delta\Theta ,\mathcal{V}_{\textrm{G$^\chi_{0}$-inv}}) \ ,
\end{equation}
where $\,B(\Theta , \delta\Theta  \, ; \, \mathcal{V}_{\textrm{G$^\chi_{0}$-inv}})\,$ accounts for the cross terms in the scalar potential (\ref{V(X)}) which are bilinear in $\,\Theta\,$ and $\,\delta\Theta\,$.

\begin{table}[t]
\begin{center}
\scalebox{0.725}{
\renewcommand{\arraystretch}{1.5}
\begin{tabular}{|cc|c|c|c|c|c|c|c|c|}\hline
 $\otimes$ & $\textbf{r}_2 \in \mathbf{133}$& $t_a{}^i$&$t_{1ij8}$&$t_{ijkl}$&$t_i{}^a$&$t_i{}^j$&$t_{aijk}$&$t_a{}^b$&$3 \,  t_a{}^a - t_i{}^i$\\
	$\textbf{r}_1 \in  \mathbf{56}$&	& $(\mathbf{6},\,\mathbf{2})_{2}$& $(\mathbf{15'},\,\mathbf{1})_{2}$& $(\mathbf{15},\,\mathbf{1})_{-2}$& $(\mathbf{6'},\,\mathbf{2})_{-2}$& $(\mathbf{35},\,\mathbf{1})_{0}$& $(\mathbf{20},\,\mathbf{2})_{0}$& $(\mathbf{1},\,\mathbf{3})_{0}$& $(\mathbf{1},\,\mathbf{1})_{0}$ \\\hline
%---------
$\Theta^{[ab]}$	&$(\mathbf{1},\,\mathbf{1})_{3}$	& $\times$ & $\times$ & $(\mathbf{15},\,\mathbf{1})_{1}$&$(\mathbf{6'},\,\mathbf{2})_{1}$&$\color{blue}{(\mathbf{35},\,\mathbf{1})_{3}}$&$(\mathbf{20},\,\mathbf{2})_{3}$&$(\mathbf{1},\,\mathbf{3})_{3}$&$\times$\\\hline
%--------------------
$\Theta_{[ij]}$	&$(\mathbf{15},\,\mathbf{1})_{1}$	&  $(\mathbf{20},\,\mathbf{2})_{3}$ & $\color{red}{(\mathbf{35},\,\mathbf{1})_{3}}$ &$\begin{matrix}(\mathbf{15'},\,\mathbf{1})_{-1}\\(\mathbf{105'},\,\mathbf{1})_{-1}\end{matrix}$ & $\begin{matrix}(\mathbf{84},\,\mathbf{2})_{-1}\\(\mathbf{6},\,\mathbf{2})_{-1}\end{matrix} $&$\begin{matrix}(\mathbf{15},\,\mathbf{1})_{1}\\(\mathbf{21},\,\mathbf{1})_{1}\\(\mathbf{105},\,\mathbf{1})_{1}\end{matrix}$&$\begin{matrix}(\mathbf{6'},\,\mathbf{2})_{1}\\(\mathbf{84'},\,\mathbf{2})_{1}\end{matrix}$&$(\mathbf{15},\,\mathbf{3})_{1}$&$(\mathbf{15},\,\mathbf{1})_{1}$\\\hline 
%--------------------
$\Theta^{[ai]}$	&$(\mathbf{6'},\,\mathbf{2})_{1}$	&  $\begin{matrix}\color{red}{(\mathbf{35},\,\mathbf{1})_{3}}\\(\mathbf{1},\,\mathbf{3})_{3}\end{matrix}$ & $(\mathbf{20},\,\mathbf{2})_{3}$ &$\begin{matrix}(\mathbf{84},\,\mathbf{2})_{-1}\\(\mathbf{6},\,\mathbf{2})_{-1}\end{matrix}$&$\begin{matrix}(\mathbf{15'},\,\mathbf{1})_{-1}\\(\mathbf{21'},\,\mathbf{1})_{-1}\end{matrix}$&$\begin{matrix}(\mathbf{6'},\,\mathbf{2})_{1}\\(\mathbf{84'},\,\mathbf{2})_{1}\end{matrix}$&$\begin{matrix}(\mathbf{105},\,\mathbf{1})_{1}\\(\mathbf{15},\,\mathbf{1})_{1}\\(\mathbf{15},\,\mathbf{3})_{1}\end{matrix}$&$(\mathbf{6'},\,\mathbf{2})_{1}$&$(\mathbf{6'},\,\mathbf{2})_{1}$\\\hline
%--------------------
$	\Theta_{[ai]}$&$(\mathbf{6},\,\mathbf{2})_{-1}$	&  $\begin{matrix}(\mathbf{15},\,\mathbf{1})_{1}\\(\mathbf{21},\,\mathbf{1})_{1}\end{matrix}$ & $\begin{matrix}(\mathbf{6'},\,\mathbf{2})_{1}\\(\mathbf{84'},\,\mathbf{2})_{1}\end{matrix}$ &$(\mathbf{20},\,\mathbf{2})_{-3}$&$\begin{matrix}(\mathbf{35},\,\mathbf{1})_{-3}\\(\mathbf{1},\,\mathbf{3})_{-3}\end{matrix}$&$\begin{matrix}(\mathbf{6},\,\mathbf{2})_{-1}\\(\mathbf{84},\,\mathbf{2})_{-1}\end{matrix}$&$\begin{matrix}(\mathbf{105'},\,\mathbf{1})_{-1}\\(\mathbf{15'},\,\mathbf{1})_{-1}\\(\mathbf{15'},\,\mathbf{3})_{-1}\end{matrix}$&$(\mathbf{6},\,\mathbf{2})_{-1}$&$(\mathbf{6},\,\mathbf{2})_{-1}$\\\hline
%--------------------
$\Theta^{[ij]}$	&$(\mathbf{15'},\,\mathbf{1})_{1}$	& $\begin{matrix}(\mathbf{84'},\,\mathbf{2})_{1}\\(\mathbf{6'},\,\mathbf{2})_{1}\end{matrix}$ & $\begin{matrix}(\mathbf{105},\,\mathbf{1})_{1}\\(\mathbf{15},\,\mathbf{1})_{1}\end{matrix}$ &$(\mathbf{35},\,\mathbf{1})_{-3}$&$(\mathbf{20},\,\mathbf{2})_{-3}$&$\begin{matrix}(\mathbf{15'},\,\mathbf{1})_{-1}\\(\mathbf{21'},\,\mathbf{1})_{-1}\\(\mathbf{105'},\,\mathbf{1})_{-1}\end{matrix}$&$\begin{matrix}(\mathbf{6},\,\mathbf{2})_{-1}\\(\mathbf{84},\,\mathbf{2})_{-1}\end{matrix}$&$(\mathbf{15'},\,\mathbf{3})_{-1}$&$(\mathbf{15'},\,\mathbf{1})_{-1}$\\\hline
%--------------------
	$\Theta_{[ab]}$&$(\mathbf{1},\,\mathbf{1})_{-3}$	&  $(\mathbf{6},\,\mathbf{2})_{-1}$ & $(\mathbf{15'},\,\mathbf{1})_{-1}$ &$\times$&$\times$&$(\mathbf{35},\,\mathbf{1})_{-3}$&$(\mathbf{20},\,\mathbf{2})_{-3}$&$(\mathbf{1},\,\mathbf{3})_{-3}$&$\times$\\\hline
\end{tabular}}
\end{center}
\caption{Contributions to the $\,\mathbf{912}\,$ originating from the tensor product $\,\mathbf{56} \times \mathbf{133}\,$. We have highlighted the three possible sources of $\,\delta\Xi \in (\textbf{35},\textbf{1})_{3}\,$. Only the blue one contributes to the second term in the bilinear $\,B(\delta\Xi ,\delta\Xi)\,$ in (\ref{V_bilinear}).}
\label{Table:56x133in912}
\end{table}

To keep the discussion as general as possible, let us introduce a generic bilinear form
\begin{equation}
\label{V_bilinear}
B(\Theta_{1} , \Theta_{2}  \, ; \, \mathcal{V}) = B(\Xi_{1} , \Xi_{2}) = \frac{1}{672}\,  (\Xi_{1})_\mathbb{M}{}^{\alpha}  \,  (\Xi_{2})_\mathbb{M}{}^{\beta}   \, (\delta_{\alpha \beta} + 7 \, \mathcal{K}_{\alpha \beta}) \ ,
\end{equation}
where we have used the alternative expression (\ref{V(Xi)}) for the scalar potential in terms of the scalar-dependent $\Xi$-tensor. We will now argue that if $\,\Xi_2 \in (\mathbf{35},\,\mathbf{1})_3 \subset\mathbf{912}\,$ (equivalently $\,\Xi_1\,$) then only the projection of $\,\Xi_1\,$ (equivalently $\,\Xi_2\,$) onto the $\,(\mathbf{35},\,\mathbf{1})_3\,$ contributes to the bilinear form (\ref{V_bilinear}). This is obvious for the first contribution to the r.h.s of (\ref{V_bilinear}) as the $\,\delta_{\alpha \beta}\,$ selects only representations of $\,\Xi_1\,$ which are present in $\,\Xi_2\,$. The analysis of the second contribution to the r.h.s of (\ref{V_bilinear}) is more subtle as terms of the form $\,(\Xi_{1})_{\mathbb{M}}{}^\alpha \, (\Xi_{2})_{\mathbb{M}}{}^{\alpha^{t}}\,$ appear by virtue of the Killing-Cartan matrix in (\ref{E7_KC}). Then, for any non-vanishing component of $\,\Xi_2 \in \textbf{r}\,$ with $\,\textbf{r}\,$ belonging to the decomposition of the $\,\textbf{912}\,$ in (\ref{branching_912-SL(6)xSL(2)}) and originating from a pair $\,(\textbf{r}_1,\,\textbf{r}_2)\,$ of irreps with $\,\textbf{r}_{1}\,$ belonging to the decomposition of the $\,\textbf{56}\,$ in (\ref{branching_56-SL(6)xSL(2)}) and $\,\textbf{r}_{2}\,$ belonging to the decomposition of the $\,\textbf{133}\,$ in  (\ref{branching_133-SL(6)xSL(2)}), only the non-vanishing components of $\,\Xi_1\,$ originating from a pair $\,(\textbf{r}_1,\,\textbf{r}_2')\,$ contribute to the second term in (\ref{V_bilinear}). In the case of $\,\Xi_2 \in (\mathbf{35},\,\mathbf{1})_3\,$ only the piece of $\,\Xi_1\,$ originating from the $\,(\textbf{r}_{1},\textbf{r}_{2}')=\big( (\textbf{1},\textbf{1})_{3} \,,\,  (\textbf{35},\textbf{1})_{0} \big)\,$ and living in the $\,(\mathbf{35},\,\mathbf{1})_3$ contributes to the second term in (\ref{V_bilinear}) as can be seen by inspection of Table~\ref{Table:56x133in912}. Parameterising the corresponding $\,T_{1,2} \in (\mathbf{35},\,\mathbf{1})_3\,$ tensors in (\ref{Xi-tensor}) in terms of two matrices $\,(\Xi_{1,2})_{i}{}^{j}\,$ as
\begin{equation}
\begin{array}{lll}
(T_{1,2})_{ij} &=&  (\Xi_{1,2})_{i}{}^k \, t_{1kj8}  - (\Xi_{1,2})_{j}{}^k \, t_{1ki8} \ , \\[2mm]
(T_{1,2})^{ai} &=& -\epsilon^{ab} \, (\Xi_{1,2})_j{}^i  \, t_b{}^j   \ , \\[2mm]
(T_{1,2})^{18} &=& (\Xi_{1,2})_i{}^j \, t_j{}^i  \ ,
\end{array}
\end{equation}
an explicit computation shows that
\begin{equation}
\label{V_bilinear_2}
B(\Xi_{1} , \Xi_{2}) = \frac{1}{192} \, \text{Tr} \left[   (\Xi_{1} + \Xi_{1}{}^t)  (\Xi_{2} + \Xi_{2}{}^t)   \right] \ .
\end{equation}
As a result, whenever $\,\Xi_{1}\,$ or $\,\Xi_{2}\,$ are $\,\mathfrak{so}(6)$-valued (\textit{i.e.} $\,\Xi_{1,2} + \Xi_{1,2}{}^t=0\,$), the bilinear (\ref{V_bilinear_2}) vanishes identically.

The analysis above has been performed in terms of scalar-dependent $\,\Xi$-tensors living in the $\,(\mathbf{35},\,\mathbf{1})_3\,$ representation. However, flat deformations were introduced in (\ref{deltaTheta-tensor}) in terms of $\,\Theta$-tensors living in the $\,(\mathbf{35},\,\mathbf{1})_3\,$ representation. Therefore, it remains to be shown that having $\,\Theta \in (\mathbf{35},\,\mathbf{1})_3\,$ implies $\,\Xi \in (\mathbf{35},\,\mathbf{1})_3\,$. Using the solvable parameterisation of the coset space $\,\textrm{E}_{7(7)}/\textrm{SU}(8)\,$ according to which scalars are associated with non-compact generators carrying non-negative SO(1,1) charge, we see from Table~\ref{Table:133x912} that the scalars acting non-trivially on $\Theta \in (\mathbf{35},\,\mathbf{1})_3\,$ separate in three families: $i)$ scalars associated with generators of $\,\textrm{SL}(6)\,$ transforming in the $\,(\mathbf{35},\,\mathbf{1})_0\,$. $ii)$ a single scalar $\,\sigma\,$ associated with the generator of $\,\textrm{SO}(1,1)\,$
\begin{equation}
t_{\textrm{SO}(1,1)} = \, \sqrt{3} \, \big( 3 \,  (t_1{}^1+t_8{}^8)-(t_2{}^2+t_3{}^3+t_4{}^4+t_5{}^5+t_6{}^6+t_7{}^7) \big) \ ,
\end{equation}
transforming in the $\,(\mathbf{1},\,\mathbf{1})_0\,$. $iii)$ scalars associated with generators of $\,\mathfrak{e}_{7(7)}\,$ transforming in the $\,(\textbf{20} , \textbf{2})_{0}\,$. Note that only scalars in the $\,(\textbf{20} , \textbf{2})_{0}\,$ are of relevance as they could generate an unwanted piece $\,\Xi \in(\textbf{20} , \textbf{2})_{+3}\,$. However, we computed that
\begin{equation}
\label{35To35}
\Xi(\Theta \, , \, \mathcal{V} ) = e^{-3 \sigma} \, \Theta \in (\mathbf{35},\,\mathbf{1})_3 \ ,
\end{equation}
provided $\,\Theta \in (\mathbf{35},\,\mathbf{1})_3\,$ and 
\begin{equation}
\label{coset-chi-inv}
[ \, \mathcal{V}  \,,\,  \chi_{i}{}^{j} \,  t_{j}{}^{i}  \,  ] = 0  \ ,
\end{equation}
with $\, \chi_{i}{}^{j} \,  t_{j}{}^{i}  \in \mathfrak{g}_{0} \subset \mathfrak{so}(6)\,$. A way of understanding (\ref{35To35}), provided (\ref{coset-chi-inv}) holds, is by alternatively thinking about the relation (\ref{Xi-tensor}) as the coset representative $\,\mathcal{V}\,$ being acted upon by the embedding tensor $\,\Theta\,$ rather than the other way around. The condition (\ref{coset-chi-inv}) severely restricts the scalar dependence of the coset representative $\,\mathcal{V}\,$ so that $\,\delta_{\chi_{i}{}^{j}} \mathcal{V}=0\,$. In particular $\,\mathcal{V}=\mathcal{V}_{\textrm{G$^\chi_{0}$-inv}}\,$ satisfies (\ref{coset-chi-inv}).

Applying the above results to the case of $\,\Theta_{1} = \Theta \in (\textbf{21},\textbf{1})_{1} \oplus (\textbf{1},\textbf{3})_{3}\,$ in (\ref{Theta_tensor}) and the flat deformation $\,{\Theta_{2}=\delta\Theta \in (\mathbf{35},\,\mathbf{1})_3}\,$ in (\ref{deltaTheta-tensor}) induced by $\mathfrak{so}(6)$-valued axions $\,\chi^{ij}=\delta^{k [i} \, \chi_{k}{}^{j]}\,$, one has that
\begin{equation}
\begin{array}{lllll}
B(\Theta , \delta\Theta  \, ; \, \mathcal{V}_{\textrm{G$^\chi_{0}$-inv}}) & = & 0 \ , \\[2mm]
B(\delta\Theta , \delta\Theta  \, ; \, \mathcal{V}_{\textrm{G$^\chi_{0}$-inv}}) &=& V( \delta\Theta  , \mathcal{V}_{\textrm{G$^\chi_{0}$-inv}}) & = &  0 \ .
\end{array}
\end{equation}
Therefore, (\ref{Vtilde_terms}) reduces to
\begin{equation}
V(\widetilde{\Theta} ,\mathcal{V}_{\textrm{G$^\chi_{0}$-inv}}) = V(\Theta ,\mathcal{V}_{\textrm{G$^\chi_{0}$-inv}}) \ ,
\end{equation}
proving the second equality in (\ref{V=V=V_appendix}).

\bibliographystyle{JHEP}
\bibliography{references}

\end{document}